# Reconstructing solar wind inhomogeneous structures from stereoscopic observations in white-light: Solar wind transients in 3D

Xiaolei Li[1,2], Yuming Wang[1,2,3,*], Rui Liu[1,2,3], Chenglong Shen[1,2,3], Quanhao Zhang[1,2], Shaoyu Lyu[1,2], Bin Zhuang[1,2], Fang Shen[4], Jiajia Liu[5], and Yutian Chi[1,3]

[1]CAS Key Laboratory of Geospace Environment, Department of Geophysics and Planetary Sciences, University of Science and Technology of China, Hefei 230026, China

[2]CAS Center for Excellence in Comparative Planetology, University of Science and Technology of China, Hefei 230026, China

[3]Mengcheng National Geophysical Observatory, University of Science and Technology of China, Mengcheng 233527, China

[4] SIGMA Weather Group, State Key Laboratory of Space Weather, Center for Space Science and Applied Research, Chinese Academy of Sciences, Beijing 100190, China

[5]Solar Physics and Space Plasma Research Center, School of Mathematics and Statistics, University of Sheffield, Sheffield S37RH, UK

*Corresponding author, ymwang@ustc.edu.cn

**Abstract** White-light images from Heliospheric Imager-1 (HI1) onboard the Solar Terrestrial Relations Observatory (STEREO) provide 2-dimensional (2D) global views of solar wind transients traveling in the inner heliosphere from two perspectives. How to retrieve the hidden three-dimensional (3D) features of the transients from these 2D images is intriguing but challenging. In our previous work (Li et al., 2018), a 'correlation-aided' method is developed to recognize the solar wind transients propagating along the Sun-Earth line based on simultaneous HI1 images from two STEREO spacecraft. Here the method is extended from the Sun-Earth line to the whole 3D space to reconstruct the solar wind transients in the common field of view of STEREO HI1 cameras. We demonstrate the capability of the method by showing the 3D shapes and propagation directions of a coronal mass ejection (CME) and three small-scale blobs during 3-4 April 2010. Comparing with some forward modeling methods, we found our method reliable in terms of the position, angular width and propagation direction. Based on our 3D reconstruction result, an angular distorted, nearly North-South oriented CME on 3 April 2010 is revealed, manifesting the complexity of a CME's 3D structure.

1. Introduction

The observation of solar wind plays a significant role on studying the impact of solar activity on the Earth, since solar wind which originates from the sun and spreads into the whole space in the Heliosphere is the main medium that transmits solar perturbations to the Earth. The solar wind can be simply classified into two categories: continuum and inhomogeneous structures. Solar wind transients are a kind of macroscopic inhomogeneous structures embedded in the solar wind continuum. With the successful launches of a series of space-borne white-light image cameras in recent decades including, e.g., the Large Angle and Spectrometric Coronagraph (LASCO,

Brueckner et al., 1995) on board Solar and Heliospheric Observatory (SOHO), coronagraphs (COR1 and COR2) and heliospheric imagers (HI1 and HI2) in the Sun Earth Connection Coronal and Heliospheric Investigation (SECCHI) suite (Howard et al., 2008) on board the Solar Terrestrial Relations Observatory (STEREO), diverse solar wind transients can be observed in a quite wide field of view (FOV) from different perspectives. Coronal mass ejections (CMEs) and blobs are representative large-scale and small-scale solar wind transients in white-light images. CMEs in the interplanetary medium (i.e., ICMEs), if fast enough, can drive an interplanetary shock, and the associated shock-sheath, the high density structure in the downstream, could be observed in white-light images (Ontiveros and Vourlidas, 2009; Maloney and Gallagher, 2011). Blobs normally originate from helmet streamers and have a slow speed, propagating along with slow solar wind (e.g., Wang et al., 1998; Sheeley et al., 1997, 2009; Rouillard et al., 2010; Plotnikov et al., 2016; Sanchez-Diaz et al., 2017).

No matter how clear the transients on a two dimensional (2D) white-light image are, it is not enough for many physical analysis without 3D information. Many methods have been developed to solve this problem. With the aid of a 3D morphological model of a transient, such as the Graduated Cylindrical Shell (GCS) model (Thernisien et al., 2006, 2009; Thernisien, 2011) for the flux rope structure of a CME and Tappin and Howard (2009) model for the leading edge of a CME, it is possible to find the best fitting result of a CME transient recorded in white-light images and therefore obtain a good estimation of its 3D geometrical and kinematic information. Transients normally leave bright or dark traces on time-elongation maps (also known as J-map, e.g., Sheeley et al., 1999; Davies et al., 2009; Sheeley et al., 2008, 2009) as they propagate outward. Based on assumptions of constant propagating velocity and certain morphology if needed, the directions and velocities of the transients, especially 'blobs', may be inferred from traces in the J-maps, and the associated methods include 'Point-P', 'Fixed-$\phi$', 'Harmonic Mean' and 'Self Similar Expansion' methods (Howard et al., 2006; Kahler and Webb, 2007; Lugaz et al., 2009; Wood et al., 2009a; Lugaz, 2010; Liu et al., 2010a, 2010b; Rouillard et al., 2010, 2011a; Davies et al., 2012; Mostl and Davies, 2013). For accelerated or decelerated transients, this method may lead to a large uncertainty. Triangulation methods were further developed to localize transient features on white-light images from dual perspectives (Howard and Tappin, 2008; Lugaz et al., 2010). It is very important to make sure the two features from the different perspectives are the same solar wind transient structure. Focusing on the boundary of the whole CME from different perspectives, Geometric Localization method (Pizzo & Biesecker, 2004; de Koning et al., 2009; Byrne et al., 2010) and its improved version, Mask Fitting method (Feng et al., 2012), can reconstruct the 3D shape of a CME's outline. For single-view polarimetric coronagraph images, polarization ratio technique (Moran and Davila, 2004; Dere et al., 2005; Moran et al., 2010; Deforest et al., 2018) makes use of the geometric dependence of the polarization of Thomson-scattered light to determine the 3D structure and positions of transients. This method is useless for unpolarized images, such as Heliopheric images of HI1 and HI2. Another method, Local Correlation Tracking (LCT) method applies a correlation analysis to establish the correspondence between pixels on two coronagraph images from two close perspectives and obtain a 3D representation of a CME (Mierla et al., 2009). For large separation angle of dual perspectives, LCT method does a poor job (Feng et al., 2013).

In our recent paper (Li et al., 2018) (hereafter referred to as Paper I), we developed a new method to use correlation analysis to recognize and locate solar wind transients along the Sun-Earth line based on STEREO/HI1 dual white-light images. The method can automatically identify the transients propagating along the Sun-Earth line from dual images. Comparing with other methods, our method is much simpler and free of any special morphological assumptions.

Now we improve the method, which we name as the CORrelation-Aided Reconstruction (CORAR) method, to retrieve 3D solar wind transients from STEREO/HI1 image data when they appear in the common FOV of the HI1 cameras. In Section 2, we introduce the STEREO/HI1 image data and the CORAR method. Section 3 analyzes the 3D region suitable for CORAR method during April 3 and 4, 2010 by applying CORAR method on synthetic HI1 images. Section 4 describes the observed solar wind transients, one CME and three blobs, during April 3 and 4, 2010. The term 'blob' here refers to any small structures travelling in the solar wind, which look like a blob and are narrow in angular width and slow in speed compared to typical CMEs. We apply the CORAR method on these transients, show the 3D reconstruction results, and compare them with those by other methods. In Section 5, we apply the CORAR method on two other CMEs on 7 and 14 August 2010 to display the influence of the propagation direction of large-scale transients on reconstruction results. Finally, we provide a summary and discussion in Section 6.

## 2. Data and Method

The white-light images we used are provided by HI1 cameras, the same as Paper I, on board the STEREO-A/B. With a FOV of 20° by 20°, pixel resolution of 1024 × 1024 and cadence of 40 minutes, STEREO/HI1 images can display the evolution of solar wind transients in detail (Harrison et al., 2008). The center of the FOV for both HI1 is at 14° in elongation and almost on the ecliptic plane (Eyles et al., 2009). With the observer-Sun-Earth separation angle of 45°-135° for both STEREO spacecraft in 2008-2012, the two HI1 can simultaneously observe many transients (see Fig. 1a). Our CORAR method is applied on two such white-light images from dual perspectives with large-enough common FOV.

The basic principle of our method is illustrated in detail in Paper I. The key idea is to place a 'baseline' with one endpoint on the Sun at first, project each HI1 images on the meridian plane in which the baseline lies, use a radial-latitude sampling box to obtain a 2D local brightness variation and then calculate the correlation coefficient (cc) along the 'baseline' (see Fig. 2 in Paper I). When a transient locates around the baseline, the brightness variations in the sampling box from two perspective will be similar, and therefore result in a high positive cc value. Then we can recognize transients around the baseline by extracting high positive cc regions. In practice, we also consider the temporal evolution of transients, and the sampling box is actually in the 3D radial-latitude-time space. By scanning the baseline in latitude and longitude, we are able to obtain the cc values in the 3D space covered by the common FOV of the two HI1 cameras and then reconstruct the 3D solar wind transients therein. The following are the detailed procedures of CORAR method.

1) **Data preprocessing.** We use the calibrated STEREO/HI-1 Level 2 images with 1 day background emissions, defined as the average of the lowest 25% of the data in a running window of 1 day on a pixel-by-pixel basis, removed (Eyles et al., 2009) as Paper I, and preprocess the image data in the same way as Paper I, which includes the shifted running-difference and 3 x 3 median filtering.

2) **Projection.** We adjust the baseline in the Heliocentric Earth Ecliptic (HEE) coordinate system in an angular range from -80° to 80° in longitude with a step of 1° and from -80° to 80° in latitude with the same step. The (0°, 0°) in longitude - latitude space corresponds to the Sun-Earth line. Therefore 161 meridian planes with HEE longitude varying from -80° to 80° are chosen to project HI1 images on.

The projection we do is just the simple optical projection. The direction, $\vec{n}$, of a pixel in HI1 image in HEE coordinate system relative to the observer, $\vec{O}$, can be obtained from the header of the STEREO/HI-1 FTS file. Assuming $\vec{P}(x, y, z)$ is the HEE coordinate of the projected point for this pixel on the meridian plane with longitude of $\phi$, we have

$$\vec{P}(x, y, z) = \vec{O} + k\vec{n} \qquad (1)$$

where k is an unknown positive number. The latitude, $\lambda$, and distance from the solar center, $D$, of the point, as the two key unknown parameters for the projection procedure, have the following relations

$$\begin{cases} x = D \cos \lambda \cos \phi \\ y = D \cos \lambda \sin \phi \\ z = D \sin \lambda \end{cases} \qquad (2)$$

By solving the equations 1 and 2, $x, y, z, k, \lambda$ and $D$ can be calculated. The derived $\lambda$ and $D$ are utilized for plotting the image projected on the meridian plane at $\phi$.

The spatial resolution of the projected images is set to 1° in latitude and about 0.2 solar radii (Rs) in radial direction. For each pixel of a projected image on a meridian plane, we combine all the projected pixels within one spatial grid by calculating their median value. Figure 2c and 2d are projected images of two HI1 images (Fig. 2a and 2b) at the same time on the meridian plane at HEE longitude of 20°. The ranges of two projected images on the same meridian plane are normally different (see the region enclosed by red/blue line in Fig. 2c/2d), and the overlapped region (i.e., the region enclosed by the red lines in Fig. 2c and 2d) of two projected images is adopted in the next procedure.

3) **Sampling and cc calculation.** Different from Paper I, we use a 3D (radial-latitude-time) running box to sample the data for the calculation of cc. The sampling box is 11-pixel (i.e., ±5°) wide in latitude and 41-pixel (i.e., ~8 Rs) long in the radial direction. It is a suitable size so that the brightness variation in the sampling box mainly depends on the density spatial variation of the transient. To enhance the density spatial variation pattern, brightness data in the sampling box is subtracted by their average value. After trial and error, we set the temporal length of the sampling box to be either 1 or 5 time-steps, and name the associated cc values as $cc_{t1}$ and $cc_{t5}$, respectively. Compared with $cc_{t1}$ with only 1 time-step, $cc_{t5}$ takes advantage of the temporal evolution of the features in the correlation analysis. The calculated cc value is then assigned to the central pixel of the sampling box.

The cc value we calculate is the linear Pearson correlation coefficient given by

$$cc = \frac{\sum_{i=1}^{n}(p_i - \bar{p})(q_i - \bar{q})}{\sqrt{\sum_{i=1}^{n}(p_i - \bar{p})^2} \sqrt{\sum_{i=1}^{n}(q_i - \bar{q})^2}} \qquad (3)$$

where p and q are two sets of data (corresponding to data in the same sampling box for STEREO-A/B) with a sample size of n, while $\bar{p}$ and $\bar{q}$ are the average values of p and q. Here in our sampling box, n=11×41×t, where t equals the number of time steps. The value of cc is between -1 and 1. The larger the value of cc is, the better is the correlation. A negative cc value is not considered a correlation here. Equation (3) suggests that the value of cc is not influenced by the densities of features but by the variation pattern of the density as long as the signal-to-noise (SNR) ratio is large enough, which will be discussed later.

4) **Finalization of the cc value.** Many factors, like the position and the SNR of a transient might affect the calculated cc value which therefore needs to be further investigated and corrected if any. The details will be given in the next section. For a sampling box with the temporal length of 1 or 5 time-steps, the corrected $cc_{t1}$ and $cc_{t5}$ have their own strengths and weaknesses. The minimum of the corrected $cc_{t1}$ and $cc_{t5}$, called $cc_{min}$, is chosen as the final cc (see Appendix A for the reason). If not specified, all the cc used hereafter is $cc_{min}$.

The 2D distribution of the final cc value in the meridian plane at longitude of 20° for then HI1 images shown in Figure 2a and 2b is displayed in Figure 2e. By putting the 2D cc distributions computed from 161 different meridian planes together, a 3D cc map is constructed with color-coded dots showing the high cc regions (Fig. 2f). Here we define a region with cc ≥ 0.5 as the high-cc region, as used in Paper I.

**3. Test of CORAR method with synthetic HI1 images**

Before apply the method to the real HI1 data, synthetic HI1 images are constructed firstly as a test to analyze the influence of the position and SNR of the features on cc and therefore the derived 3D structures.

3.1 Synthetic running-difference HI1 image with a spherical blob

Assuming that all the transient features in HI1 images could be approximated as a combination of small blobs, we create a synthetic running-difference HI1 image as a sum of a simulated spherical blob as the signal and a selected background image as the noise (see Figure 3a, 3b). The selected background images are real STEREO-A/B running-difference HI1 images at the moment when the solar wind was extreme quiet without any notable transient. The spherical blob is given by

$$n_e(\overrightarrow{R_b} + \vec{r}) = \begin{cases} n_0 \cdot \frac{1}{2}\left(1 + \cos\left(\pi \frac{|\vec{r}|}{r_0}\right)\right), & |\vec{r}| \leq r_0 \\ 0, & |\vec{r}| > r_0 \end{cases} \quad (4)$$

where $n_e$ is the electron number density, $\overrightarrow{R_b}$ is the coordinate of the blob center, $\vec{r}$ is the coordinate of any one point relative to the blob center, $r_0$ is the radius of the blob and $n_0$ is the number density at the blob center. The STEREO-A/B are assumed at the positions at 00:09 UT on 4 April, and the Thomson scattering brightness of the blob is calculated (by following Howard and Tappin 2009; Howard and DeForest 2012) as

$$B(\varepsilon, \alpha) = B_\odot \frac{\pi \sigma_t r_\odot^2}{R^2 \sin^2 \varepsilon} \int_0^{+\infty} (1 - \cos^4 \chi) n_e(z, \varepsilon, \alpha) dz \quad (5)$$

where $B_\odot$ is the solar surface brightness, $r_\odot$ is the length of Rs, $\sigma_t$ is the area of Thomson scattering cross-section, R is the distance from solar center to the observer, $\chi$ is the angle of scatter, $\varepsilon$ is the Sun-observer-scattering point angle (also called elongation angle), $\alpha$ is an azimuthal angular coordinate and z is the distance from the scattering point to the observer. After being converted the simulated blob to the brightness, it is further subtracted by itself with a radial shift of -0.2 $r_0$ before being inserted into the selected background image. To make sure the density pattern has a similar size of our sampling box (with a radial size of 8 Rs), the $r_0$ is set to be 5 Rs. The value of $n_0$ is set to be $10^3$, $10^4$ and $10^5$ $cm^{-3}$, standing for low, middle and high density, respectively. We put the synthetic spherical blob at 170 different positions within in the common FOV of the two HI1 cameras (see small dots in Fig. 6). The 170 positions spread from about 30 to 90 Rs in distance, -30° to 30° in latitude and -50° to 50° in longitude in HEE coordinate system (see Fig. 6).

3.2 Influence of SNR on cc and the method of correction

When applying CORAR method on real or synthetic HI1 images, the SNR of the data in a sampling box during sampling process varies a lot for transients with different intensity level. The definition of this SNR and its calculation method are introduced in Appendix B. Please note that the signal talked in this study is pure signal from the transients, not including noise. To figure out how this SNR influences the value of cc, we try to do variable separation on cc about SNR

$$\text{cc} = cc_R \cdot cc_M \qquad (6)$$

where $cc_M$ is a function of SNR alone while $cc_R$ is independent of SNR. The value of $cc_M$ is bounded between 0 and 1, corresponding to zero and infinite SNR, respectively, so that $cc_R$ is equal to cc of two data sets without any noise (i.e. infinite SNR).

In our two synthetic HI1 images, signal patterns of the two sampled data sets should be the same if the simulated blob is captured by the two sampling boxes, and their $cc_R$ is 1 and $cc_M$ is just equal to cc. But the SNR in images will change with the density and position of the simulated blob. Thus, by putting the blob at different positions with different density, we can obtain the relation between $cc_M$ and SNR.

The SNR of the two image sets, $SNR_A$ and $SNR_B$, and their corresponding $cc_M$ for the blob at 170 different positions with different density are displayed in Figure 4a. The value of the $cc_M$ increases from 0 to 1 along with the increase of $SNR_A$ and $SNR_B$. When $SNR_A$ and $SNR_B$ are both greater than about 10, the $cc_M$ is nearly 1. To simplify the two binary function of $cc_M(SNR_A, SNR_B)$, we define the total SNR as

$$TSNR \equiv \sqrt{SNR_A^2 + SNR_B^2} \ . \qquad (7)$$

Based on the $cc_M$-TSNR plot in Figure 4b, we find that $cc_M$ could be related to TSNR by the following equation

$$cc_M = 1 - \frac{1}{1+\frac{TSNR^2}{2}} \qquad (8)$$

As shown by the red curve in Figure 4b. Clearly, $cc_M$ is less than 0.3 if TSNR is smaller than 1, and is close to 1 if TSNR is larger than 3. And there is no clear dependence of $cc_M$ on the position of the simulated blob, suggesting the SNR is the main factor affecting the calculated cc values.

Based on the above analysis, we may infer that the corresponding SNR is too small to recognize real features if cc at a position is less than 0.3. For such cases, we do not correct their cc values and treat them uncorrelated. For other cases, cc is corrected by $cc/cc_M$.

3.3 Validation of the reconstruction of simulated blobs

By applying the CORAR method on all the synthetic HI1 images, 3 × 170 3D cc maps are constructed (see Fig. 3d is an example). Because 1 time-step is the only option of temporal length for synthetic HI1 images, the final cc in 3D cc maps here is $cc_{t1}$. According to the principle stated in Section 2, the high-cc region in 3D cc map should match the position of its corresponding transient. In Figure 3, blob has $n_0$ of $10^4\ cm^{-3}$, and locates at 10° in HEE latitude, -19° in HEE longitude and 50 Rs away from the sun. The final cc value in the blob is greater than 0.5 and it rapidly falls down to around 0 outside the blob (shown in Fig. 3f-3h). The high-cc region and this synthetic spherical blob overlap with each other very well (see Fig. 3e).

We then compare the position of the high-cc region with the simulated blob for all the 170 positions, and find that not all the recognized high-cc regions well overlap with the corresponding blobs. The central position of a high-cc region is given by its cc-weighted average distance, latitude and

longitude

$$\langle D \rangle = \frac{\sum_i cc_i D_i}{\sum_i cc_i} \tag{9}$$

$$\langle \lambda \rangle = \frac{\sum_i cc_i \lambda_i}{\sum_i cc_i} \tag{10}$$

$$\langle \phi \rangle = \frac{\sum_i cc_i \phi_i}{\sum_i cc_i} \tag{11}$$

where $D_i$, $\lambda_i$ and $\phi_i$ are distance from the solar center, latitude and longitude of any point in the high-cc region. Then we calculated the position deviation between the high cc region and the corresponding blob $\Delta D = \langle D \rangle - D_b$, $\Delta \lambda = \langle \lambda \rangle - \lambda_b$, $\Delta \phi = \langle \phi \rangle - \phi_b$, in which $D_b$, $\lambda_b$ and $\phi_b$ are the distance, latitude and longitude of the blob center, respectively.

Figure 5a -5c display the results. The deviation $\Delta D$, $\Delta \lambda$ or $\Delta \phi$ increases with the increasing distance of the blob, and the deviation for the low density blobs is much larger than that for the middle and high density blobs. At the distance less than about 60 Rs, the absolute mean distance or angular deviation is less than 1 Rs or 1° with its standard deviation less than 1 Rs or 1°, which is small enough for locating especially compared with that over 60 Rs. As the latitude or longitude of the blob increases, the deviation keeps low value without obvious tendency of increasing. Compared with the distance, latitude and longitude play little influence on the locating accuracy.

The size of the high-cc region is also a key factor evaluating the accuracy of the derived location, which is expected to be not too larger than the simulated blob. The cc-weighted standard deviations of distance, latitude and longitude of the high-cc region, are calculated to represent the half size of the reconstructed structure in the three dimensions

$$R_{ccD} = \sqrt{\frac{\sum_i cc_i (D_i - \langle D \rangle)^2}{\sum_i cc_i}} \tag{12}$$

$$\alpha_{cc\lambda} = \sqrt{\frac{\sum_i cc_i (\lambda_i - \langle \lambda \rangle)^2}{\sum_i cc_i}} \tag{13}$$

$$\alpha_{cc\phi} = \sqrt{\frac{\sum_i cc_i (\phi_i - \langle \phi \rangle)^2}{\sum_i cc_i}} \tag{14}$$

The half size of a simulated blob in distance, latitude and longitude is $r_0$, $\alpha_b$ and $\alpha_b$, respectively, where we define $\alpha_b \equiv r_0/D_b$ as the half angular width of the blob. Figure 5d-5f show the results of $R_{ccD}/r_0$, $\alpha_{cc\lambda}/\alpha_b$ and $\alpha_{cc\phi}/\alpha_b$.

As figure 5e and 5f show, the lower density, the less value of $R_{ccD}/r_0$ and $\alpha_{cc\lambda}/\alpha_b$. Especially for the low density (the blue symbols), the reconstructed structure mostly is smaller than the simulated blob, mainly because of the low SNR near the edge of the blob. At the distance greater than 60 Rs, $R_{ccD}/r_0$ and $\alpha_{cc\lambda}/\alpha_b$ are normally larger than that within 60 Rs especially for large density blobs (see left column of Fig. 5d and 5e), suggesting a lower locating accuracy beyond 60 Rs. Similar to $\Delta D$, $\Delta \lambda$ and $\Delta \phi$, $R_{ccD}/r_0$ and $\alpha_{cc\lambda}/\alpha_b$ change little as latitude or longitude changes, suggesting a week influence of angular position on the locating accuracy.

The behavior in longitude, i.e., $\alpha_{cc\phi}/\alpha_b$, is a little bit different, which increases with the blob's coordinate along x axis' direction (i.e., Sun-Earth direction) in HEE coordinate system (see Fig. 5f). We think it is mainly influenced by 'collinear effect' (see Paper I) that leads to an abnormal high-cc region widely spreading along the connecting line of two spacecraft if there is a transient near the connecting line. Here we extend the concept of 'collinear effect' to 'coplanar effect', which means that in 3D the abnormal high-cc region will extend to a plane, we call 'co-plane', passing through the spacecraft and perpendicular to the Sun-spacecraft plane. As shown in Figure 5f, when transient propagates close to the 'co-plane' (the green line), $\alpha_{cc\phi}/\alpha_b$ rises rapidly. At the HEE x coordinate of less than about 60 Rs, $\alpha_{cc\phi}/\alpha_b$ is typically less than 1.2 suggesting a weak influence of 'coplanar effect'.

Based on the analysis above, we set the following criteria to evaluate the goodness of the reconstruction in our test

$$\begin{cases} \Delta D < 2 \text{ Rs} \\ \Delta \lambda < 2° \\ \Delta \phi < 2° \\ \frac{R_{ccD}}{r_0} < 1.2 \\ \frac{\alpha_{cc\lambda}}{\alpha_b} < 1.2 \\ \frac{\alpha_{cc\phi}}{\alpha_b} < 1.2 \end{cases} \qquad (15)$$

The color-coded dots in Figure 6 summarize the results. Those in cyan indicate the positions satisfying all the criteria for all the three densities. The dots in white only simultaneously meet the first three criteria but not the last three criteria. The orange dots indicate the places unsuitable for the reconstruction. It could be found that about 70% of the 170 positions are suitable for the reconstruction, and only less than 10% of the positions are not suitable if the density is not too low. The blobs closer to the Sun can be reconstructed better than those far away from the Sun. Especially below 60 Rs, almost all of the blobs are well reconstructed.

In summary, the test suggests that in most regions of the common FOV of the two STEREO/HI1 cameras (i.e., below 60 Rs, corresponding to the elongation angle less than 20°), small-scale transients can be accurately recognized and located by the CORAR method. The situation of a large-scale transient must be much more complicated than that of a single small-scale transient. Here, as the first-order approximation, we simply treat a large-scale transient as a combination of many small-scale transients, and correct the cc value by the same way.

**4. The solar wind transients during 3 and 4 April, 2010**

4.1 Observations

We apply the CORAR method to the HI1 data on 3 and 4 April 2010 to reconstruct solar wind transients, when a large-scale transient, CME, and three small-scale blob-like transients were propagating in the HI1 FOV (see the animation M1.mp4 in supporting materials). The separation angle between STEREO-A/B was 138° (Fig. 1a).

The 3 blob-like small transients, labeled as Blob 1-3, originated and propagated along the corona streamer in the FOV of STEREO-COR2 and first appeared in the HI1 FOV at 12:09 UT on 2 April, 04:49 UT and 10:09 UT on 4 April, respectively, and kept a clear white front and dark back lasting for over 12 hours in the running-difference images (see Fig. 7a, 7b, 7e and 7f). As they propagated in the HI1 FOV, they expanded and distorted from initially simple blobs to more diffused structures. Since they are narrow in angular width and slow in speed compared with typical CMEs, we consider them as blobs rather than CMEs.

After the first blob, a CME propagated outward radially (Liewer et al., 2015) into HI1 FOV at 13:29 on 3 April (see Fig. 7c and 7d). The CME drove a bright ribbon, mostly apparent in front of the CME's northern part and stuck to the CME's southern part. Considering that there was a shock ahead of the CME in the in-situ observations near the Earth (Möstl et al., 2010), we deduce the bright ribbon as the CME driven shock. The density structure pattern inside the CME looks complicated, which includes two bright ribbons extending over a large latitudinal range named as "driver gas" structures by Rouillard et al. (2011b). This CME was associated with a B7.4 class X-ray flare from the active region 11059 (S25W03 in Heliocentric Earth equatorial coordinate system) beginning at 09:04 UT on April 3, 2010 (Liu et al., 2011; Xie et al., 2011) and a pre-existing loop which rose around 08:40 UT or earlier (Liu et al., 2011) and erupted around 09:16 UT (Xie et al., 2011). Before the CME was seen by HI1 at 12:09 UT, it had been observed by LASCO and SECCHI COR2 since 09:54 UT. Kinematic analysis based on J-maps constructed by COR2, HI1 and HI2 observations suggests that the bright front of this CME was rapidly accelerated to about 1000 km s$^{-1}$ and then slightly decelerated after 12:00 UT (Liu et al., 2011; Wood et al., 2011; Xie et al., 2011). In situ observations by ACE and Wind spacecraft at the L1 point near the Earth recorded the corresponding interplanetary shock and ICME event during 5 and 6 April 2010, confirming that the CME was Earth-directed. Möstl et al. (2010) couldn't find a typical magnetic cloud (MC) structure in this ICME and suggested it was because the spacecraft traveled through the north flank instead of the center of the ICME.

4.2 The cc distribution in 3D

The 3D cc maps constructed from the HI1 images during 3 and 4 April 2010 by the CORAR method are shown in Figure 8 (an animation, M2.mp4, could be found in the supporting materials which is produced by Python packages, Mayavi (Ramachandran and Varoquaux, 2010)). The three blobs are all recognized as indicated by the high-cc regions in the top and bottom panels. The selected high-cc regions corresponding to the CME event are as complicated as its 2D pattern in HI1 images. We can identify the shock front ahead of the northern part of the CME like a hat on the other high-cc regions. The constructed 3D shape of the CME is quite different from any simple flux-rope morphology model, indicating the complexity of such a fast and large CME in interplanetary space.

4.3 Validation of the reconstruction of the CME

To validate the 3D reconstruction result of the CME, especially the inferred propagation direction and angular width, we use three other CME models to fit the CME and compare the results with ours. First, the GCS model is used. Considering the significant distortion of this CME in the HI1 images, we choose STEREO/SECCHI COR2 images and SOHO-LASCO images to do the reconstruction. The GCS model involves 3 positioning parameters: θ, φ –the Stonyhurst latitude and longitude of the source region, γ –the tilt angle of the source region neutral line, and 3 geometric parameters: H –the height of the leading edge, κ -the aspect ratio, δ -the half edge-on angular width (Thernisien et al., 2006; Thernisien 2011). We obtain the best-fitting estimation of the parameters θ, φ, γ, κ, δ and transform them to HEE coordinate system to give the direction and angular extent of the CME. The values of the fitting parameters are listed in Table 1, suggesting that the CME propagated toward -19° in latitude and 4° in longitude with its front shape nearly East-West (E-W) extended (see dark blue line in Fig. 9b).

Different models derive different angular extent of the CME. Xie et al. (2011) use Krall's and St Cyr's (2006) flux rope model which they name as KS06 to fit this CME. They calculate the best-fit $\omega_{edge}$ and $\omega_{broad}$, the widths of the CME from edge-on and face-on and orientation parameters: θ, φ -the Stonyhurst latitude and longitude, γ –the tilt angle are also converted to HEE coordinate system (see Tab. 1), and the angular extent is indicated by blue lines in Fig. 9b. They believe that its front is almost North-South (N-S) extended.

Different from the GCS and KS06 models which only fit SECCHI COR2 and LASCO images, Wood et al. (2011) use their empirical 3D reconstruction model (Wood et al., 2009b) to fit SECCHI/HI1 and HI2 data. They provide two fitting results with the tilt angle γ, of 10° (i.e., almost E-W) and -80° (i.e., almost N-S). We name this model as ER model and derive the best-fit parameters: θ, φ, γ, $\omega_{edge}$ and $\omega_{broad}$ (similar definition to KS06, see Tab.1) in HEE coordinate system and give its latitude and longitude extent (see green and grass green lines in Fig. 9b) as well.

We also display this CME recognized by our method in the following way. At any given time, we calculate the 'thickness' of high-cc regions, which is defined as the number of high-cc points along every radial direction. These numbers are normalized by their maximum value to make sure that the value of thickness fall into the range of 0 and 1. From 16:09 UT to 21:29 UT on 3 April 2010, the main transient on HI1 images was only this CME. Before 16:09 UT, the whole CME structure had not entered the common FOV of HI1, while after 21:29 UT the CME propagated too close to the connecting line of STEREO-A/B and the 'coplanar effect' becomes significant large. Thus, we select the time of 16:49 UT to show the thickness of the reconstructed CME.

As Figure 9b shows, most of the 'thick' regions locate within the border of the best-fit scope by the three models on this CME. It means that our method correctly reconstructs the direction and angular extent of the CME. At latitudes greater than 10°, there is a 'thick' region out of the scope of most models. The reason is that this region corresponds to the shock front, which is not considered in those models.

The 'thickness' map in Figure 9b also show the longitude of the CME on the ecliptic plane is about 5°±10°. By using geometric triangulation method, Liu et al. (2011) inferred the longitude of this CME of around 10°. With 'Fixed-φ' method, this longitude were found to be 9° by Möstl et al. (2010), 2° by Wood et al. (2011) and 3°±4° by Rollett et al. (2012). Using 'Harmonic Mean' method, Möstl et al. (2010) and Xie et al. (2011) derived this longitude to be -5° and -8°. Our result is consistent with their estimates. However, Möstl et al. (2014) calculated the longitude to be -19° with 'Self Similar Expansion' method, while Temmer et al. (2011) and Rollett et al. (2012) derived the longitude of -25°±10° by 'Harmonic Mean' method, which have a substantial difference from our result. This is because a high velocity would lead to a large error in the CME propagation direction (Lugaz and Kintner, 2013; Möstl et al., 2014).

4.4 Validation of the reconstruction of blobs

Since the shape of a 'blob' is simple compared to the CME, we use the ice-cream cone model (Fisher & Munro, 1984) to reconstruct them and compare the results with ours. The ice-cream cone model is composed of a ball which we call the ice-cream ball and a circular cone tangent to the ball with a conic node on the solar surface. Thernisien (2011) showed that the GCS model becomes equivalent to the ice-cream cone model when its parameter δ equals 0. We apply the ice-cream cone model simplified from the GCS model on the three blobs to estimate their four key parameters: θ, φ -the Stonyhurst latitude and longitude of the cone axis, κ -the aspect ratio equal to the sine value of half-angle of cone and H -the Heliocentric height of the leading edge, and convert them to HEE coordinate system (see Tab.1). The ice-cream cone model results are shown in Figure 9a and 9c-9d for the three blobs, respectively. Most of the 'thick' regions derived by the CORAR method are within the scope of the blobs in the ice-cream cone model as marked by the skyblue circles, which means that our method also correctly reconstruct the direction and angular extent of the blobs.

Figure 10 shows the evolution of the reconstructed blobs by our method and the ice-cream cone model. The selected periods for the three blobs are 00:09 UT to 12:09 UT on 3 April 2010, 09:29 UT to 20:09 UT on 3 April 2010 and 14:09 UT to 23:29 UT on 3 April 2010, respectively. In these periods, the blobs fully entered the region suitable for the CORAR reconstruction as discussed in Section 3.3. In our method, the cc-weighted average distance, latitude, longitude ($\langle D \rangle$, $\langle \lambda \rangle$ and $\langle \phi \rangle$, in HEE coordinate system) is used as the center of the blobs (the purple diamonds in Fig.10), and the cc-weighted standard deviation of distance, latitude and longitude ($R_{ccD}$, $R_{cc\lambda}$ and $R_{cc\phi}$) as the uncertainty or size of the blobs (the purple error bars). The positions and the longitudinal and latitudinal extent of the blobs derived from the ice-cream cone model are represented as the black solid lines and shadow regions, respectively. The radial extent of the blobs is estimated as the radius of the modeled ice-cream ball.

It is found from Figure 10, the 3D positions of the three blobs by our method match the results from the ice-cream cone model fairly well, suggesting again that our method can reconstruct transients like 'blobs' well. There are small jitters in the reconstructed longitude and latitude of the blobs by our CORAR method, but still within the range estimated by the cone model. The radial/angular deviation between the results from two method at the distance less than 60 Rs is normally less than 2 Rs/2°, which is consistent with the results concluded in Section 3.3.

Besides, we also plot the parameters of the derived blobs in Figure 5 (marked as purple dots) for comparison with the synthetic results. But for the observed blobs, we do not have their real 3D geometrical information. Thus, we use the results of ice-cream cone model as the referential geometrical information of the blobs, i.e., $D_b = H$, $\lambda_b = \theta$, $\phi_b = \varphi$, $\alpha_b = arcsin(\kappa)$ and $r_0 = H \cdot \alpha_b$. Then we calculate $\Delta D$, $\Delta \lambda$, $\Delta \phi$, $R_{ccD}/r_0$, $\alpha_{cc\lambda}/\alpha_b$ and $\alpha_{cc\phi}/\alpha_b$ by using equation (9)-(14) for Blob 1-3 at all the times. It could be found that most of the data points of the observed blobs meet the criteria we set in Section 3.3, which is consistent with the results in test. In Figure 5f, $\alpha_{cc\phi}/\alpha_b$ of the observed blobs is generally larger than the synthetic values though most of the data points below 1.2, implying that the 'coplanar effect' is probably more significant in real images than that in the synthetic images.

## 5. The CME events on 7 August 2010 and 14 August 2010

The CME on 3 and 4 April 2010 propagated too close to the central plane between the two spacecraft about which they are symmetric. Although in Section 3, we have shown that the angular position or the propagation longitude has little effect on the reconstruction accuracy for small-scale blobs. But CMEs are large-scale structures, we evaluate the effect of propagation longitude by studying two additional CMEs which propagated along the direction between the Earth and STEREO-A/B (Shen et al., 2014, see Tab. 1 for details). One CME propagated along the direction of (36°E, 6°S) in the HEEQ system according to the GCS modeling result, and first appeared in two HI1 images at around 21:29 UT on 7 August 2010. The other CME propagated along the direction of (42°W, 11°S), and first appeared in two HI1 images at 12:49 UT on 14 August 2010. The separation angle between the two STEREO spacecraft was 150° and 152° for the two CMEs, respectively (see Figure 1b and 1c). We follow exactly the previous steps to construct the two CMEs and create the 'thickness' maps as shown in Figure 11. The times of the two snapshots are 4:49 UT on 8 August 2010 and 17:29 UT on 14 August 2010, respectively, when both CMEs had fully entered the common FOV of the two HI cameras and were well below 60 Rs, close to which the 'coplanar effect' becomes significant based on the test with the synthetic HI1 images in Section 3. The GCS model is applied to the two CMEs for comparison. It is found that the 'thick' regions derived by our method roughly match the angular extents estimated by the GCS model, but a deviation between the reconstructed structure and the GCS modeled structure is evident. For the two CMEs, the CORAR method gives a direction closer to the central plane between the two spacecraft by about 20 degrees no matter if the CME propagated toward east or west. Though the CORAR is applied to HI1 images and GCS to COR2 and LASCO images, we may still conclude that the 3D reconstruction of large-scale transients are biased toward the central plane between the two spacecraft. This bias might be able to be corrected if it depends on the deviation from the central plane, which could be further studied in the future.

## 6. Summary and Discussion

We have developed a new method, CORAR, to recognize and locate solar wind transients based on two simultaneous STEREO-A/B HI1 images. This method does not presume any morphology of transients, and can be run in an automated way. Through the application of the CORAR method on the simulated blobs, we find that within a large region of the common FOV of the two HI1 cameras (particularly below the distance of 60 Rs), the blobs can be recognized and located accurately. We further test the CORAR method with the HI1 image data on 3-4 April 2010, and retrieve the 3D positional and geometrical information of the solar wind transients including a CME and three blobs. Our reconstruction results match the results from other forward modeling fairly well, proving the reliability of the CORAR method.

Some weaknesses also exist in the CORAR method. For example, the 'coplanar effect' can lead to a very low longitudinal locating precision for transients near the 'co-plane', and the reconstruction of the large-scale transients has the bias toward the central plane between the two spacecraft. Besides, the CORAR method has a potential problem to mistakenly treat two different parts of a wide transient as the same one just like triangulation methods (Howard et al., 2012) because we both use the simple optical projection. Hence, for observers on the ecliptic plane, the CORAR method locates small transients like blobs better than large-scale transients like CMEs, and locates N-S oriented CMEs better than E-W oriented CMEs in consideration of their longitudinal thickness.

Our method has many similarities with the LCT method (Mierla et al., 2009), such as the use of correlation analysis to find the possible 3D position of similar patterns on two images. The main difference between the two methods is the preprocessing of images for correlation analysis. For LCT method, the images are rectified at positions given by the epipolar line coordinate and the horizontal coordinate (Mierla et al., 2009) but in our method the two images are projected on the meridian plane first and then resampled in latitude-radial space. So the LCT method focuses on the local patterns of transients on the plane perpendicular to each camera's line of sight while our CORAR method focuses on the local patterns on the meridian plane where the solar wind transients propagate, which is much closer to the reality. Furthermore, when the separation angle between the two spacecraft is large, say larger than 90°, the LCT method performs poorly (Feng et al., 2013), but our method performs well.

The 'thickness' maps in Figures 9b and 11 show that the shapes of the CMEs are irregular. It should be clarified that the CORAR method is designed to recognize any inhomogeneous density features from imaging data. Thus, a reconstructed structure may include all the disturbed features associated to the CMEs, e.g., the CME-driven shock and shock sheath. However, even with these multiple components considered, the reconstructed CMEs still do not look like a simple and coherent flux rope as assumed in traditional empirical CME's flux rope models, but look like a distorted one. So far, our CORAR method shows promising potential in reconstructing the 3D position and geometry of solar wind transients in the common FOV of two HI1 cameras. In the future, we will further derive 3D velocity maps and plasma density maps of solar wind transients based on the 3D cc maps. The CORAR method can be applied on white-light image data not only from STEREO-A/B but also any two spacecraft with large enough common FOV of their heliospheric imagers or coronagraphs. In spite of the data missing of STEREO-B since 2013, it is still possible to apply the CORAR method from the joint observations of STEREO-A, SOHO, even the Parker Solar Probe and the Solar Orbiter on solar wind transients. We also expect to apply our method to the future space mission with white-light imagers, like L4/L5 mission and Solar Ring missions (Wang et al., 2020).

**Acknowledgments** The STEREO/SECCHI data are produced by a consortium of NRL (USA), RAL (UK), LMSAL (USA), GSFC (USA), MPS (Germany), CSL (Belgium), IOTA (France), and IAS (France). The SOHO/LASCO data are produced by a consortium of the Naval Research Laboratory (USA), Max-Planck-Institut für Aeronomie (Germany), Laboratoire d'Astronomie (France), and the University of Birmingham (UK). The SECCHI data presented in this paper were obtained from STEREO Science Center(https://stereo-ssc.nascom.nasa.gov/data/ins_data/secchi_hi/L2). The SOHO data presented in this paper were obtained from https://seal.nascom.nasa.gov/ archive/soho/private/data/processed/lasco/level_05/. We acknowledge the use of them. We thank the anonymous referees for valuable comments and suggestions. This work is supported by the grants from the Strategic Priority Program of the Chinese Academy of Sciences (XDB41000000 and XDA15017300), the NSFC (41842037, 41804161, 41774178, 41574165, 41761134088, 4184200073 and 41750110481) and the fundamental research funds for the central universities (WK2080000077).

**Appendix A**
After trial and error, we set the temporal length of the sampling box to be either 1 or 5 time-steps, and name the associated corrected cc values as $cc_{t1}$ and $cc_{t5}$, respectively. It could be proven that both $cc_{t1}$ and $cc_{t5}$ have their own merits. On the occasion when there is only noise in the sampling box at the first four time steps and a transient enters at the fifth, $cc_{t5}$ will result in a false high-cc value but $cc_{t1}$ won't (see Fig. Ab). On another occasion when two transients with different projective propagation speed look similar by coincidence at several times in the sampling box, we will mistakenly treat them as the same transient according to $cc_{t1}$ but $cc_{t5}$ will lead to a proper low cc value (see Fig. Ac). Thus, we choose the minimum of $cc_{t1}$ and $cc_{t5}$ as the final cc, called $cc_{min}$. As illustrated by Figure A and Table 2, using $cc_{min}$ we can mostly reduce the false high-cc pixels and recognize real features.

**Appendix B. Calculation of the signal-to-noise ratio (SNR) in the sampling box**
After the data preprocessing of CORAR method, the photon noise from background starfield and F coronal is mostly removed and its level is controlled to less than 10-15 Mean Solar Brightness (MSB). During sampling process, in a sampling box, we have a data set $B_i = S_i + N_i$, in which B is the intensity recorded in the running difference image, S and N are the pure signal part and the noise part, respectively, and the subscript i=1, 2, …, n indicates the points in the sampling box. The

SNR is defined as the ratio of the pure signal level to the noise level. Here the pure signal level, SL, and the noise level, NL, are defined as the standard deviation of $S_i$ and $N_i$

$$SL = \sqrt{\frac{1}{n}\sum_i (S_i - \langle S_i \rangle)^2}$$

$$NL = \sqrt{\frac{1}{n}\sum_i (N_i - \langle N_i \rangle)^2}$$

On the other hand, the standard deviation of $B_i$ is

$$SD = \sqrt{\frac{1}{n}\sum_i (B_i - \langle B_i \rangle)^2} = \sqrt{\frac{1}{n}\sum_i (S_i + N_i - \langle S_i \rangle - \langle N_i \rangle)^2}$$

$$= \sqrt{\frac{1}{n}\sum_i (S_i - \langle S_i \rangle)^2 + \frac{1}{n}\sum_i (N_i - \langle N_i \rangle)^2 + \frac{2}{n}\sum (S_i - \langle S_i \rangle)(N_i - \langle N_i \rangle)}$$

In consideration of the uncorrelation between signal and noise, the sum of the coupling terms $\sum (S_i - \langle S_i \rangle)(N_i - \langle N_i \rangle)$ should be 0, and therefore $SD = \sqrt{SL^2 + NL^2}$. We then get

$$SNR \equiv \frac{SL}{NL} = \sqrt{(\frac{SD}{NL})^2 - 1} \ .$$

For synthetic HI1 images, *NL* in a sampling box is accurately calculated based on the selected background images introduced in Section 3.1. For any real HI1 images, there is no way to separate noises from observed signals, and *NL* in a sampling box cannot be obtained. We just use 6×10$^{-16}$ MSB as *NL*, which is the mean value of *NL* in all sampling boxes calculated based on the selected background images.

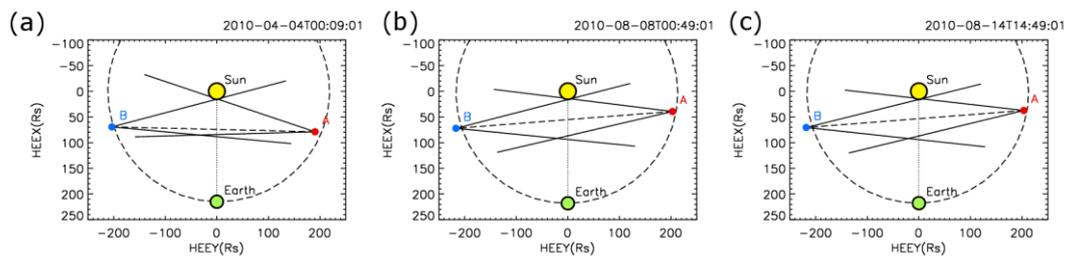

Figure 1. Positions of STEREO-A/B, Earth and the Sun on the ecliptic plane at (a) 00:09:01 UT on 4 April 2010, (b) 00:49:01 UT on 8 August 2010 and (c) 14:49:01 on 14 August 2010. The black dashed circle line represents the Earth's orbit. Black solid lines indicate the angular extents of FOVs of both HI1 cameras in the ecliptic plane. Black dash line is the connecting line of STEREO-A/B, where the collinear-effect happens. The Sun-Earth line is shown as a dotted line.

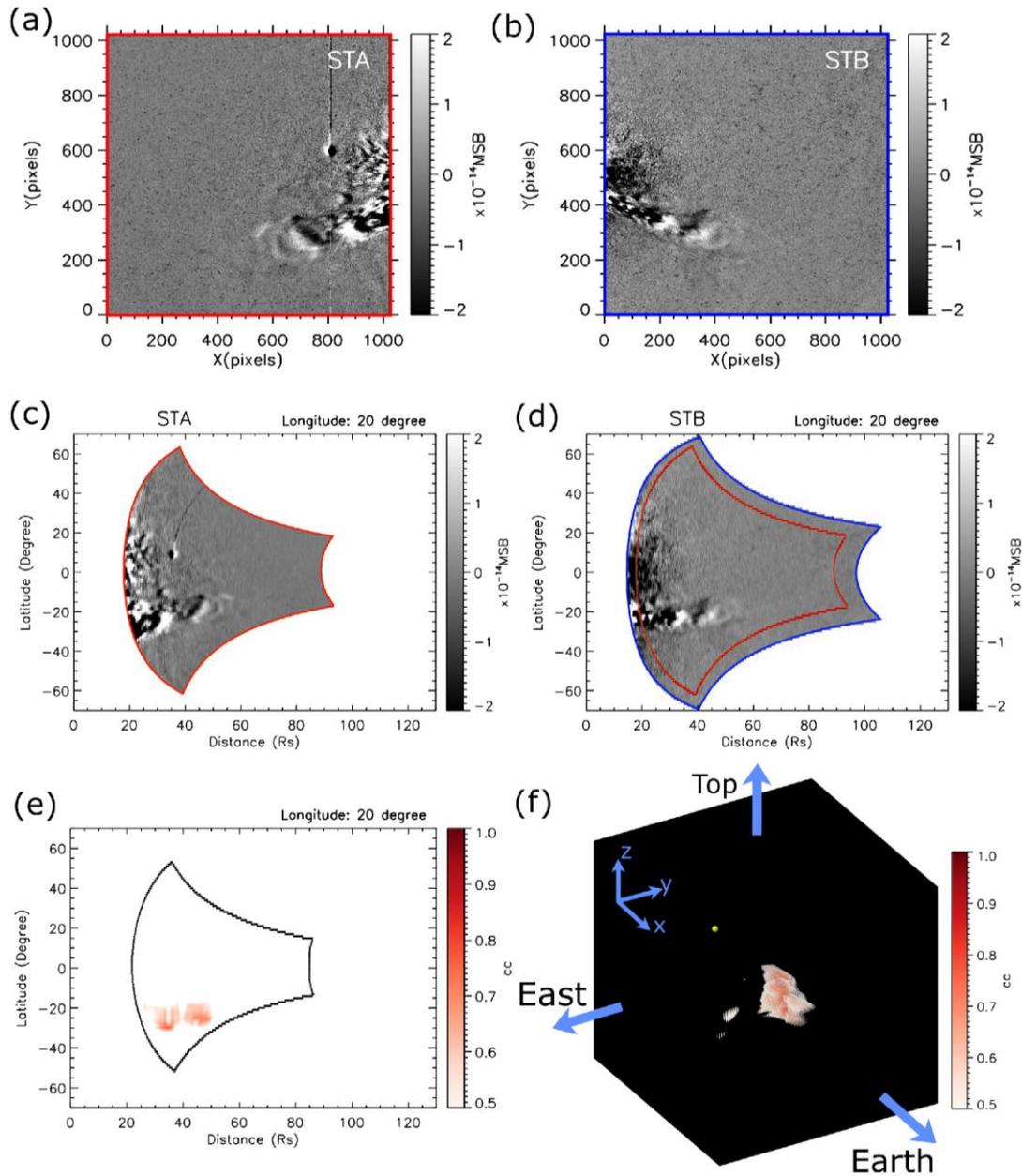

Figure 2. The example of procedures of CORAR method. Panels (a) and (b): The running-difference HI1 images for STEREO-A/B at the same time. Panels (c) and (d): The projected images of (a) and (b) on the meridian plane at longitude of 20°. The red/blue line in Panel (c)/(d) displays the border of FOV for STEREO-A/B as that in Panel (a)/(b). The red line in Panel (d) marks the overlapped region with Panel (c). Panel (e): The 2D distribution of the final cc value in the meridian plane. The black line encloses the region where the cc value is calculated. It is slightly smaller than the overlapped region of the two projected images, as the sampling box has a size (see Sec.2). Panel (f): The 3D cc map constructed by putting total 161 2D cc maps at different meridian planes together. The color-coded dots indicate the high cc (i.e., cc ≥ 0.5) regions. The larger yellow dot indicates the Sun.

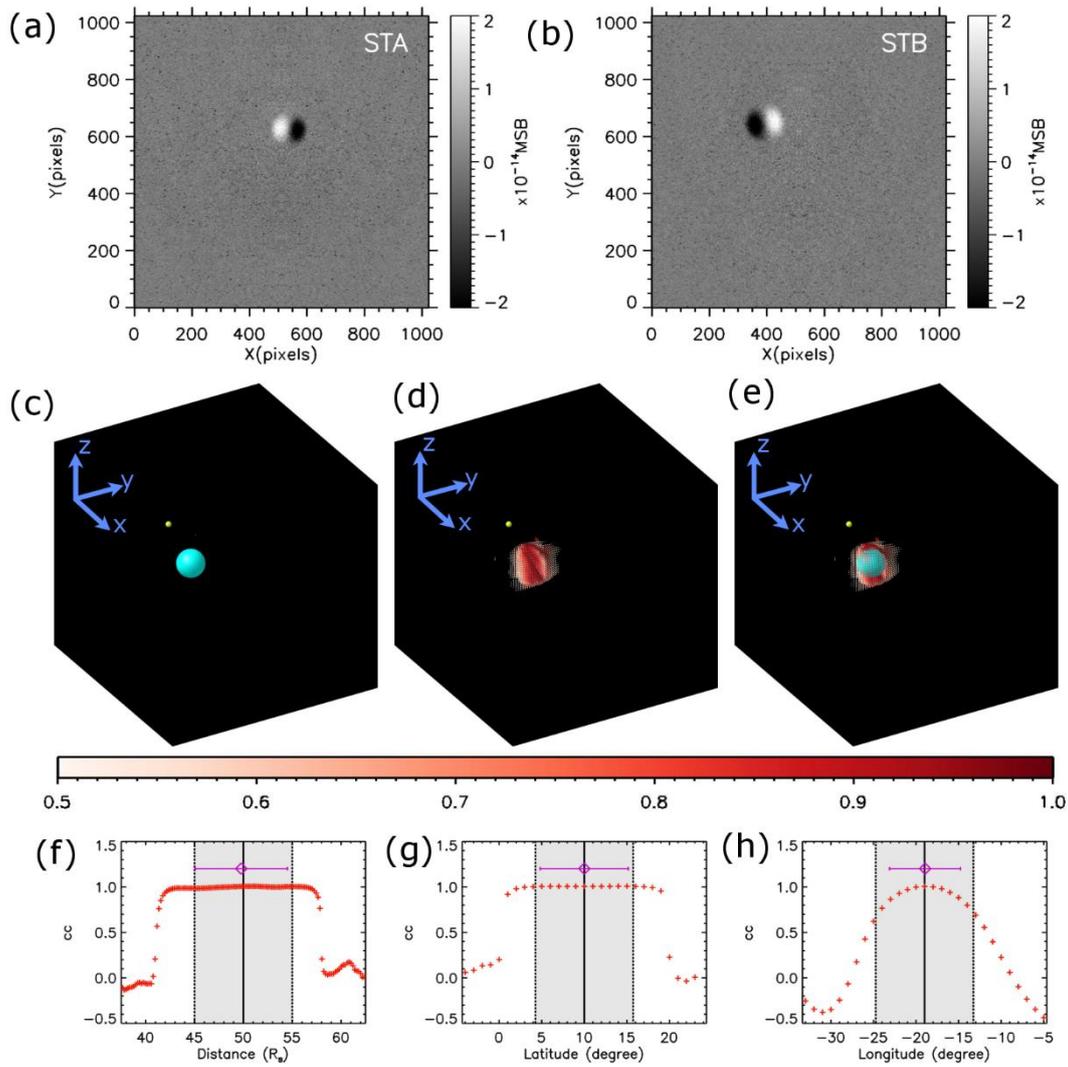

Figure 3. Panels (a-b): One example of synthetic HI1 images with a synthetic spherical blob. The simulated blob has a radius of 5 Rs and central density of $10^4$ cm$^{-3}$ at 10°, -19°, 50 Rs in HEE latitude-longitude-distance coordinate system. Panels (c-e): show the 3D position of the blob, the reconstructed high-cc region and the overlap of them, respectively in the cube of -10 Rs ≤ x ≤ 90 Rs, -50 Rs ≤ y ≤ 50 Rs and -50 Rs ≤ z ≤ 50 Rs. The large yellow dot is the Sun. Panels (f-h): The the scatter plot of the dots in the 3D cc map with the same latitude and longitude/distance and longtude/distance and latitude as the center of the blob. The black vertical solid lines plot the distance, the HEE latitude and longitude of the center of the blob. The shadow regions show their ranges for the blob. The purple diamonds show the cc-weighted average distance, latitude and longitude in the high-cc region (i.e. cc ≥ 0.5). The purple error bars show the corresponding cc-weighted standard deviation.

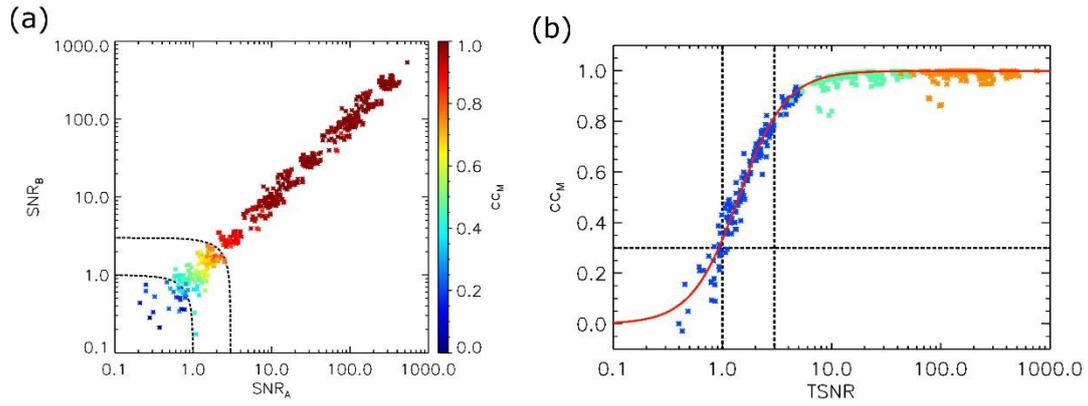

Figure 4. The relationship between $cc_M$ and SNR in the result of our test involving 170 different positions and 3 different densities. Panel (a): The value of $cc_M$ as a function of $SNR_A$ and $SNR_B$. The value of $cc_M$ is scaled by the color bar on the right. The two dash lines give the contours of $TSNR$ = 1 and 3, respectively. Panel (b): The scatter plot of the $cc_M$ versus $TSNR$. The blue, green and orange asterisks represent the data for blobs of low, middle and high density (i.e., $10^3$, $10^4$ and $10^5$ $cm^{-3}$), respectively. The red curve follows Equation 8.

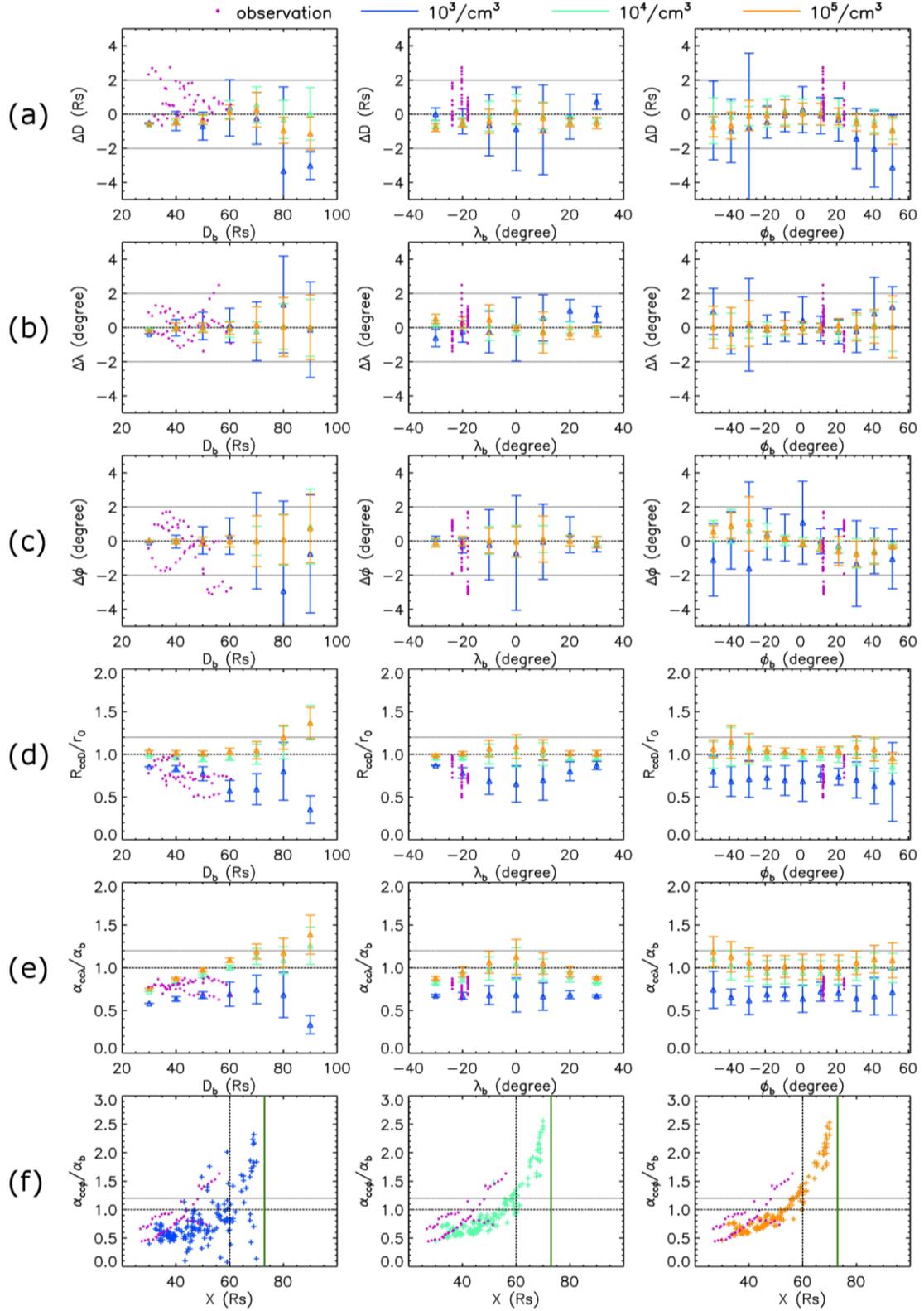

Figure 5. The calculated geometrical parameters of high cc region for the test with central electron number density of $10^3$ (blue), $10^4$ (green) and $10^5$ (orange) $cm^{-3}$, respectively, and for the three observed blobs on 3 and 4 April 2010 (purple dots). Panels (a)-(c): the radial/latitudinal/longitudinal deviation between the high-cc region and its corresponding blob as a function of the distance, latitude and longitude of the blob, respectively. Panels (d) and (e): the relative radial and latitudinal size of high-cc region as a function of the distance, latitude and longitude of the blob, respectively. Panel (f): the relative longitudinal size of high-cc region as a function of the distance. In Panels (a)-(e), the synthetic data points are grouped in bins, and triangles and error bars give the mean values and standard deviations of the blobs in the bins. In Panel (f): the synthetic data points are presented with the plus sign, and the green vertical line marks the position of 'collinear plane', which is at $x_{cp} = 73$ Rs in our test. The black vertical dotted line is at 60 Rs, roughly dividing the high and low $\frac{\alpha_{cc\phi}}{\alpha_b}$ regions.

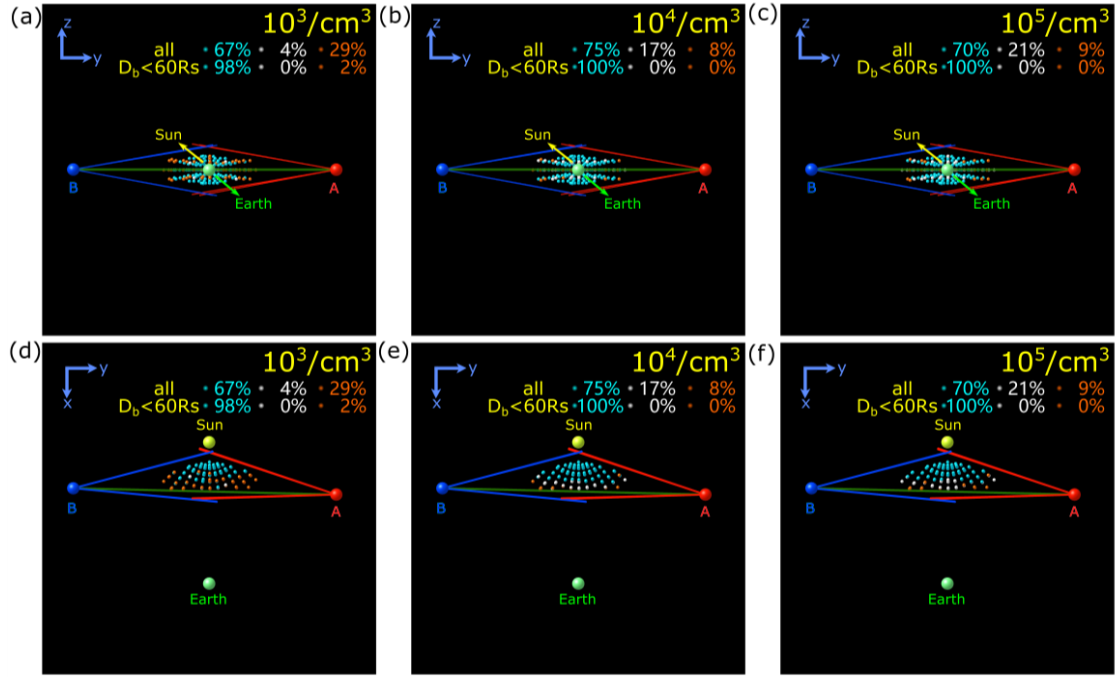

Figure 6. The distribution of the goodness of the reconstruction based on the tests for 170 positions and three different densities. From the left to right column, the test results for low, middle, high-density blobs are represented. The upper panel shows the view along the Sun-Earth line, and the lower panel shows the view from the top. The cyan dots indicate the places where the simulated blobs can be well reconstructed, i.e., meeting the criteria (15). The white dots the places where the simulated blobs can be recognized but their sizes may not be correctly retrieved, i.e., only meeting the first three criteria in (15). The orange dots the places where the simulated blobs are not well reconstructed. The percentage of each kind of reconstructed blobs for all the 170 positions or for the positions below 60 Rs is labeld on the panels. The large yellow, light green, red and blue balls are the Sun, the Earth, STEREO-A and STEREO-B. The red/blue lines denote the FOV of STEREO-A/B HI1. The green line is the connecting line of STEREO-A/B.

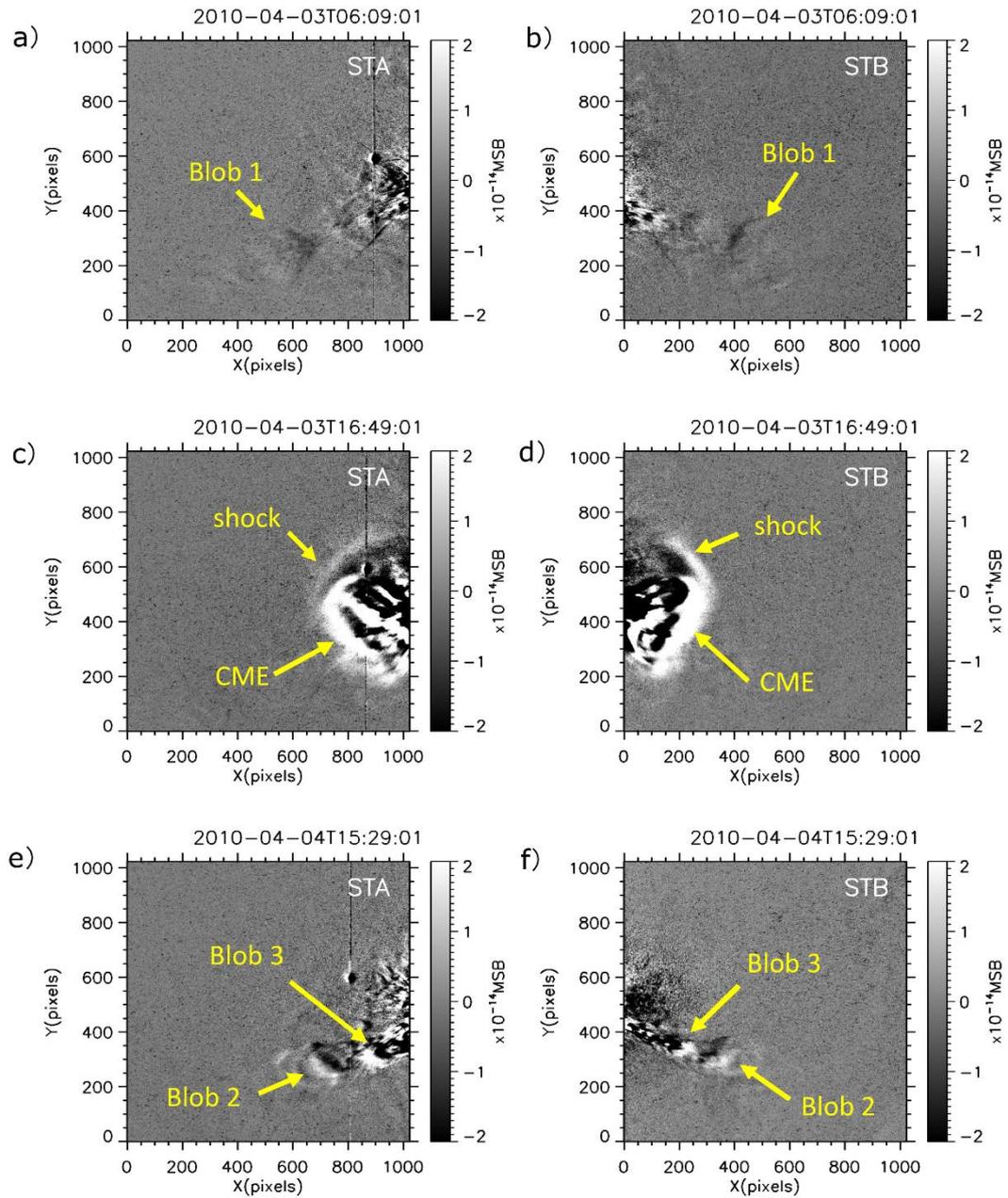

Figure 7. The HI1 running-difference images for (Panels c-d) the CME and (Panels a-b, e-f) the three blobs of interest. The left/right column is for STEREO-A/B. An animation, M1.mp4, is provided in the supporting materials.

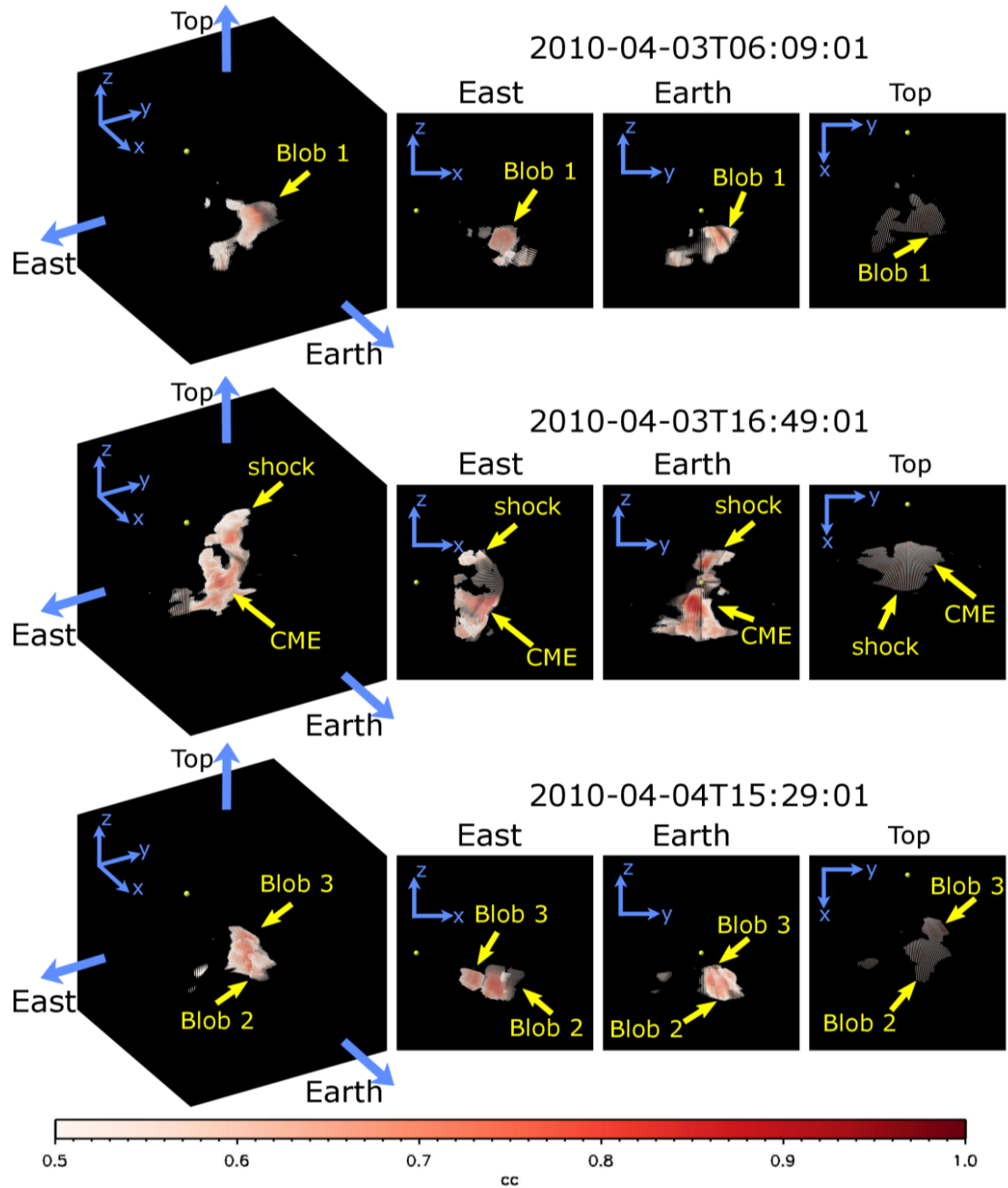

Figure 8. The high-cc regions in the cube with x-range of (-10, 90) Rs, y-range of (-50, 50) Rs, z-range of (-50, 50) Rs in HEE coordinates. The four columns each display the high-cc regions observed from a normal angle of view, the east (along the positive y axis direction in HEE coordinates), the Earth (along the negative x axis direction in HEE coordinates) and the top/north (along the negative z axis direction in HEE coordinates) from the left to the right. The three rows each display the high-cc regions at 06:09 UT on 3 April 2010 (corresponding to Fig 2a and 2b), 16:49 UT on 3 April 2010 (corresponding to Fig 2c and 2d) and 15:29 UT on 4 April 2010 (corresponding to Fig 2e and 2f). The large yellow dot is the Sun. The tiny dots display the 3D position of points with $cc_{min}$ greater than 0.5. An animation, M2.mp4, is provided in the supporting materials

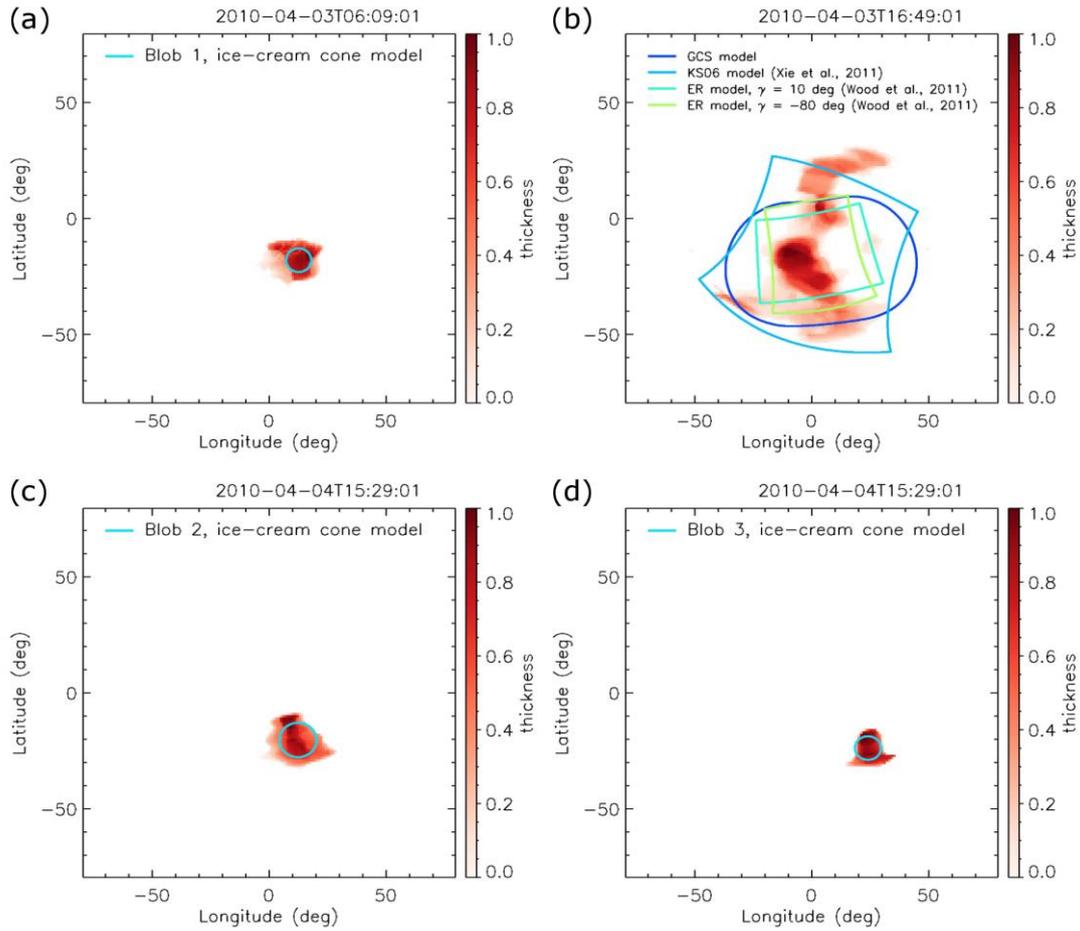

Figure 9. The 'thickness' of high-cc regions (see Section 4.2 for details) showing the angular extent of the reconstructed transients. From Panel (a) to (d), the blob 1/CME/blob 2/blob 3 are represented, respectively, at the times as same as those in figure 7 and 8. The skyblue lines in panels except (b) display the border of the best-fit of ice-cream cone model. The color-coded lines in Panel (b) display the border of the CME in different models (see main text for details).

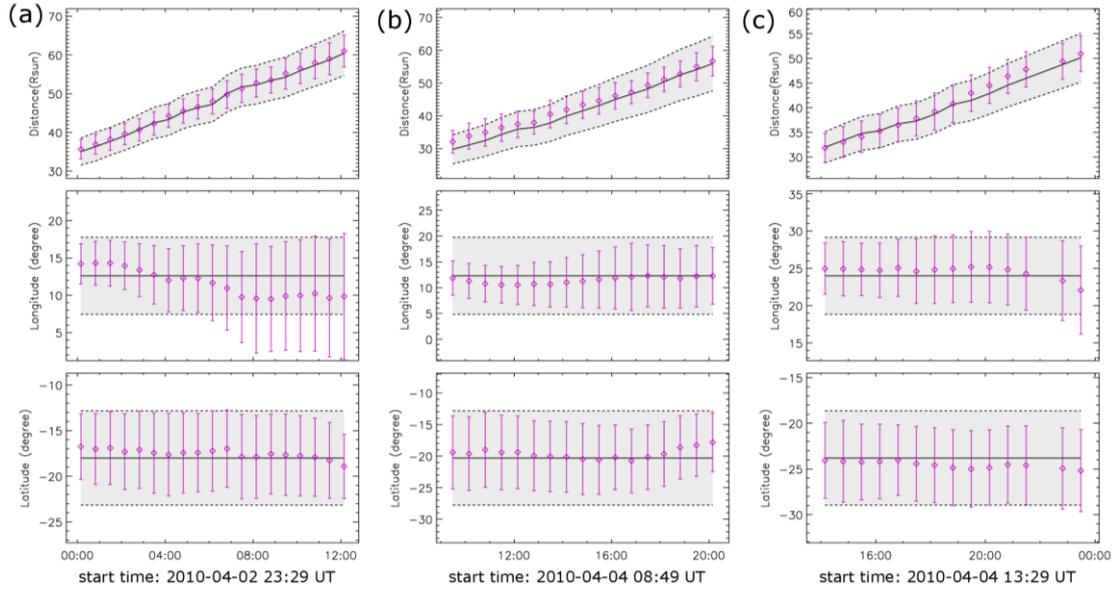

Figure 10. The locations and sizes of Blob 1-3 (columns a-c) inferred from the CORAR method (purple diamonds and error bars) and the ice-cream cone model (black lines and shadow regions). From the top to bottom panels, the distance from the solar center, the HEE latitude and longitude of reconstructed Blob 1-3 are plotted. The black lines and the shadow regions show the center and the range of the fitted ice-cream ball. The purple diamonds and error bars show the cc-weighted center and standard deviation of the high-cc region (i.e. cc $\geq$ 0.5) for Blob 1-3.

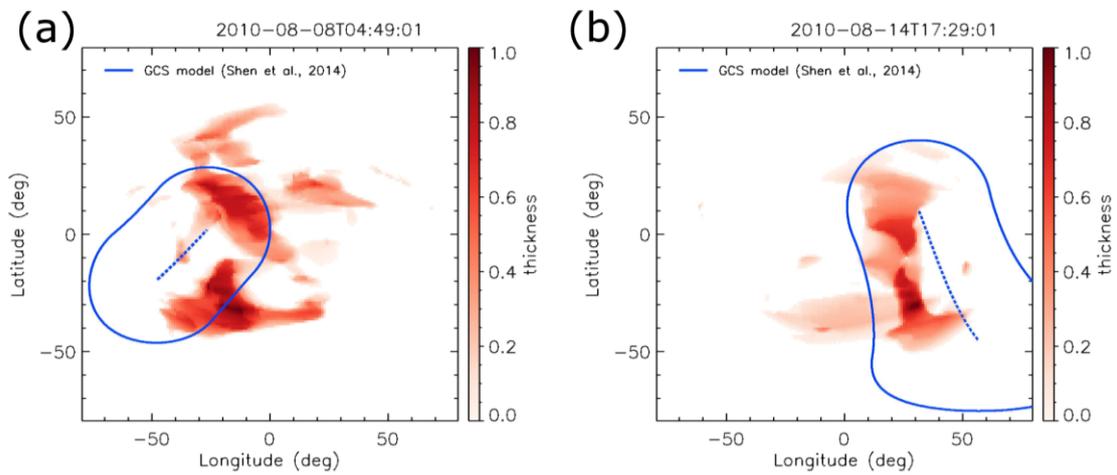

Figure 11. The 'thickness' of the two other CMEs. The solid and dashed blue lines denote the border and axis of the best-fit GCS model.

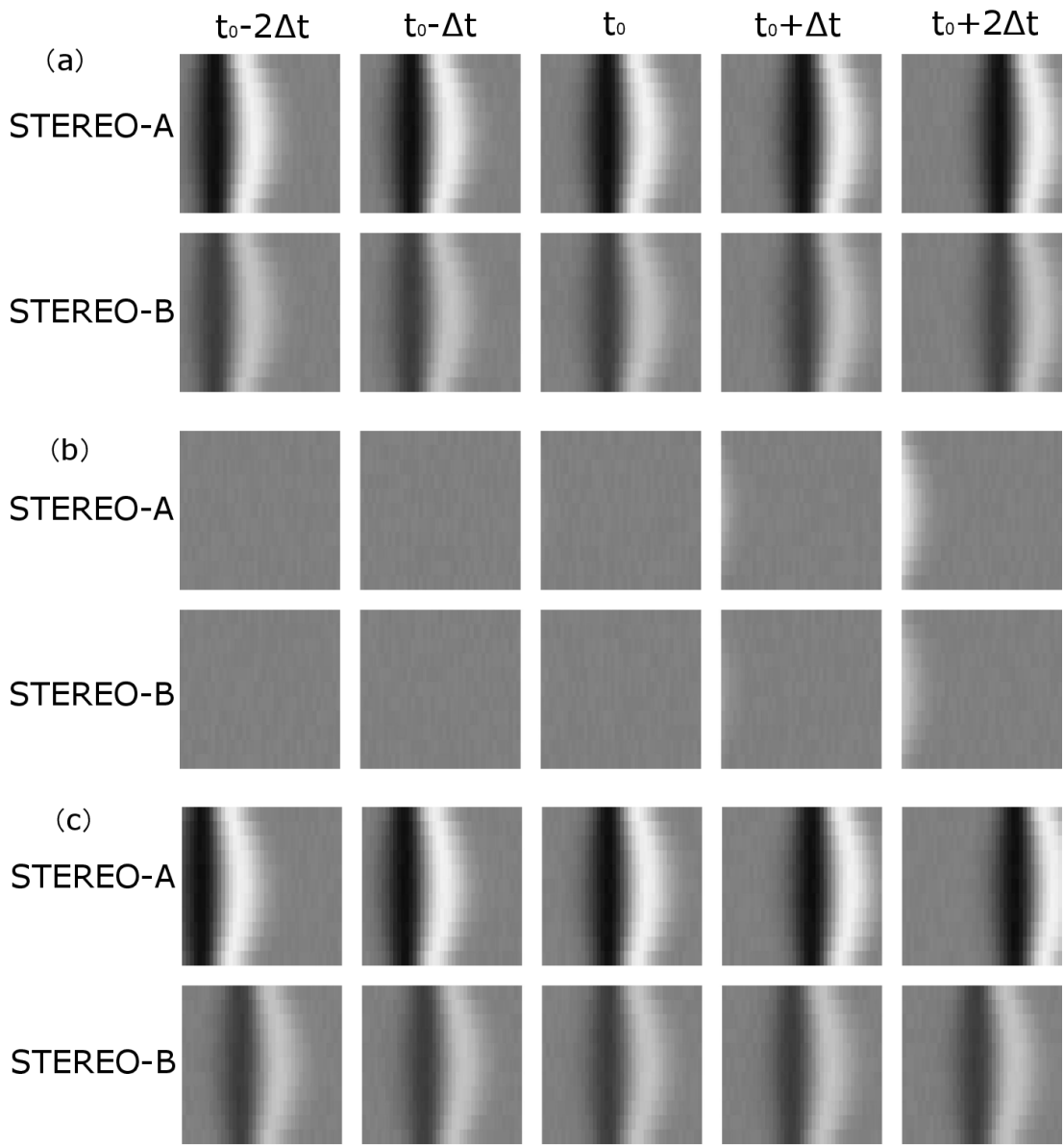

Figure A. Three groups of synthetic white-light STEREO/HI1 image data in 3D (time-latitude-radial) sampling box. $t_0$ is the current time and $\Delta t$ is the temporal sampling interval, normally 40 minutes for STEREO/HI1. The sampling box is 11-pixel wide in latitude (i.e., vertical direction) and 41-pixel wide in the radial direction (i.e., horizontal direction). The temporal length of the sampling box is either 1 (i.e., only at $t_0$) or 5 time-steps (i.e., from $t_0 - 2\Delta t$ to $t_0 + 2\Delta t$). (a) Group 1: a normal transient on the base line. (b) Group 2: there is only noise in the sampling box at the first four moments and a transient enters at the fifth moment. (c) Group 3: two transients with different projective propagation speed look similar by coincidence at several moments in the sampling box.

Table 1: The parameters of the transients in different models in HEE coordinate system.

| Date | Transient | Model | φ(°) | θ(°) | γ(°) | κ | δ(°) | $\omega_{edge}$(°) | $\omega_{broad}$(°) |
|---|---|---|---|---|---|---|---|---|---|
| 2010-04-03 | Blob 1 | Ice-cream cone | 12.6 | -18.0 | / | 0.09 | / | 10.4 | 10.4 |
| 2010-04-03 | CME | GCS | 4.3 | -18.9 | 6.4 | 0.45 | 12.0 | 53.5 | 77.5 |
| | | KS06 | 1.8 | -16.7 | 66.7 | / | / | 64.0 | 73.7 |
| | | ER | 1.5 | -16.0 | 10 | / | / | 35.9 | 47.4 |
| | | ER | 1.5 | -16.0 | -80 | / | / | 35.9 | 47.4 |
| 2010-04-04 | Blob 2 | Ice-cream cone | 12.3 | -20.3 | / | 0.13 | / | 15.0 | 15.0 |
| 2010-04-04 | Blob 3 | Ice-cream cone | 24.0 | -23.8 | / | 0.09 | / | 10.4 | 15.0 |
| 2010-08-08 | CME | The GCS | -37.0 | -9.0 | 46.0 | 0.45 | 15.0 | 53.5 | 83.5 |
| 2010-08-14 | CME | The GCS | 42.0 | -18.0 | -69.0 | 0.50 | 30.0 | 60.0 | 120.0 |

Table 2: The value of $cc_{t1}$, $cc_{t5}$ and $cc_{min}$ of three groups of synthetic data in Figure A.

| Group | The same transient observed at $t_0$? | $cc_{t1}$ | $cc_{t5}$ | $cc_{min}$=min($cc_{t1}$, $cc_{t5}$) |
|---|---|---|---|---|
| 1 | Yes | 1.00 | 1.00 | 1.00 |
| 2 | No | 0.26 | 0.99 | 0.26 |
| 3 | No | 1.00 | 0.17 | 0.17 |